\documentclass{aa}

\usepackage[varg]{txfonts}
\usepackage{enumerate}
\usepackage{hyperref}
\usepackage{graphicx}
\usepackage{color}
\usepackage[dvipsnames]{xcolor}
\usepackage{natbib}
\usepackage{longtable}
\usepackage{caption}
\usepackage{amssymb}
\usepackage{overpic}
\usepackage{float}
\usepackage{pdfpages}
\usepackage[morefloats=70]{morefloats}

\usepackage{rotating}

\graphicspath{{./figures/}}

\def \Lya{\ensuremath{\mathrm{Ly}\alpha}\ }
\def \Lyan{\ensuremath{\mathrm{Ly}\alpha}}
\def \Lyac{\ensuremath{\mathrm{Ly}\alpha},\ }
\def \Lyap{\ensuremath{\mathrm{Ly}\alpha}.\ }
\def \kms {\ifmmode  \,\rm km\,s^{-1} \else $\,\rm km\,s^{-1}  $ \fi }
\def \Lsun {\ifmmode L_{\odot} \else $L_{\odot}$ \fi}
\def \Msun {\ifmmode M_{\odot} \else $M_{\odot}$\fi}

\begin{document}

\title{Discovery of a faint, star-forming, multiply lensed, Lyman-$\alpha$ blob}

\author{G.~B.~Caminha       \inst{\ref{unife}}\thanks{e-mail address: \href{mailto:gbcaminha@fe.infn.it}{gbcaminha@fe.infn.it}} \and
        W.~Karman           \inst{\ref{Kapteyn}}\thanks{e-mail address: \href{mailto:karman@astro.rug.nl}{karman@astro.rug.nl}}  \and
        P.~Rosati           \inst{\ref{unife}}                          \and
        K.~I.~Caputi        \inst{\ref{Kapteyn}}                        \and
        F.~Arrigoni~Battaia \inst{\ref{eso}}                            \and
        I.~Balestra         \inst{\ref{obs_munich},\,\ref{inaftrieste}} \and
        C.~Grillo           \inst{\ref{dark}}                           \and
        A.~Mercurio         \inst{\ref{inafcapo}}                       \and
        M.~Nonino           \inst{\ref{inaftrieste}}                    \and
        E.~Vanzella         \inst{\ref{inafbologna}}                    
        }
\institute{
Dipartimento di Fisica e Scienze della Terra, Universit\`a degli Studi di Ferrara, Via Saragat 1, I-44122 Ferrara, Italy\label{unife}\and
Kapteyn Astronomical Institute, University of Groningen, Postbus 800, 9700 AV Groningen, The Netherlands \label{Kapteyn} \and
European Southern Observatory, Karl-Schwarzschild-Str.2, D-85748 Garching b. München, Germany \label{eso}\and
University Observatory Munich, Scheinerstrasse 1, 81679 Munich, Germany\label{obs_munich}\and
INAF - Osservatorio Astronomico di Trieste, via G. B. Tiepolo 11, I-34143, Trieste, Italy\label{inaftrieste}\and
Dark Cosmology Centre, Niels Bohr Institute, University of Copenhagen, Juliane Maries Vej 30, DK-2100 Copenhagen, Denmark\label{dark}\and
INAF - Osservatorio Astronomico di Capodimonte, Via Moiariello 16, I-80131 Napoli, Italy\label{inafcapo}\and
INAF - Osservatorio Astronomico di Bologna, Via Ranzani 1, I- 40127 Bologna, Italy\label{inafbologna}
}

\abstract
{
We report the discovery of a multiply lensed Lyman-$\alpha$ blob (LAB) behind the galaxy cluster AS1063 using the Multi Unit Spectroscopic Explorer (MUSE) on the Very Large Telescope (VLT). The background source is at $z=$ 3.117 and is intrinsically faint compared to almost all previously reported LABs.
We used our highly precise strong lensing model to reconstruct the source properties, and we find an intrinsic luminosity of $L_{\Lya}$=$1.9\times10^{42}$ erg s$^{-1}$, extending to 33 kpc. We find that the LAB is associated with a group of galaxies, and possibly a protocluster, in agreement with previous studies that find LABs in overdensities. In addition to Lyman-$\alpha$ (Ly$\alpha$) emission, we find \ion{C}{IV}, \ion{He}{II}, and
\ion{O}{III}] ultraviolet (UV) emission lines arising from the centre of the nebula. We used the compactness of these lines in combination with the line ratios to conclude that the \Lya nebula is likely powered by embedded star formation. Resonant scattering of the \Lya photons then produces the extended shape of the emission. 
Thanks to the combined power of MUSE and strong gravitational lensing, we are now able to probe the circumgalatic medium of sub-$L_{*}$ galaxies at $z\approx 3$.
}

\titlerunning{Discovery of a faint, star-forming, multiply lensed, Lyman-$\alpha$ blob by MUSE}

\authorrunning{G.~B.~Caminha et al.}

\maketitle

\section{Introduction}
\label{sec:introduction}

Understanding important processes involved in galaxy evolution, 
such as accretion and feedback, is crucial to
comprehending the medium surrounding galaxies at high redshift
In this light, understanding the nature
of the so-called Lyman-$\alpha$ blobs (LABs) provides key elements of the
conditions of the circumgalactic medium (CGM) at high redshift. These LABs are extended nebulas of
tens to hundreds of kpc with high Lyman-$\alpha$ (Ly$\alpha$) luminosities of 10$^{42}$ to 10$^{45}$ 
erg s$^{-1}$ cm$^{-2}$ \citep[e.g.][]{Heckman1991a,Steidel2000,Matsuda2011,Cantalupo2014}.

While it has become clear that most LABs reside
in over dense regions, such as the protocluster SSA22 where a large number of 
LABs have been found \citep[e.g.][]{Yamada2012}, the mechanism responsible for extended \Lya emission is still a matter of debate. 
Although active galactic nuclei (AGNe) are often shown at the centres of LABs \citep[e.g.][]{Geach2009,Overzier2013}
and several quasi stellar objects (QSO) and high-z radio galaxies (HzRG) show extended \Lya emission, other studies find  
photoionization by star formation \citep[e.g][]{Bridge2013,Ao2015,Patricio2015} as a cause for the extended emission. Others suggest that the absence of any clear ioniziation source indicates that cold accretion provides the extended \Lya emission (e.g. \citealt{Nilsson2006,Saito2006,Saito2008}, but see \citealt{Prescott2015} for the discovery of a hidden AGN in this sample).

Currently, four different scenarios are proposed for ionization of the blobs 
\citep[see][and references therein]{Arrigoni2015a}. First, an AGN could photoionize the medium, and it
has been well established that for some LABs this is the dominant factor \citep[e.g.][]{Geach2009,Overzier2013}. 
Second, shocks powered by galaxy-scale outflows could interact with the
CGM and  the intergalactic medium (IGM) producing the large amount of radiation observed \citep[e.g.][]{Mori2006}. 
The third proposed mechanism is cooling radiation through gravitational collapse 
\citep[e.g.][]{Yang2006,Dijkstra2009,Faucher2010}, which occurs when extended
metal-poor collisionally-excited gas cools down by \Lya emission. 
Last, if a source, such as a star formation (SF) burst or an AGN, is embedded in a 
large reservoir of neutral gas, resonant scattering of \Lya photons 
can extend the originally compact \Lya emission \citep[][]{Dijkstra2008,Hayes2011,Steidel2011}.

It has been shown that a single mechanism cannot explain the 
origin of the LABs for all sources. It is therefore important for our 
understanding of galaxy formation to determine which
mechanism is responsible in the different classes of objects. One method
to distinguish between the mechanisms is to look at the extent of the 
\ion{C}{IV} and \ion{He}{II} emission. Extended \ion{C}{IV} and \ion{He}{II} emission is
expected when photoionization or shocks power the \Lya nebula, while
only extended \ion{He}{II} emission is expected in case of gravitational
cooling.
Because \ion{He}{II} is not a resonant line, no extended \ion{He}{II} emission is expected in case of a resonant scattering.
Although \ion{C}{IV} is resonant, it will only be extended for resonant scattering if the surrounding medium is optically thick to \ion{C}{IV} \citep[see][]{2014ApJ...796..140P}.

Although extended \ion{C}{IV} and \ion{He}{II} emission are detected around
HzRGs \citep[][]{Villar2003,Villar2007,Humphrey2006,Gulberg2015,Swinbank2015}, detections of extended
\ion{C}{IV} and \ion{He}{II} are elusive around QSOs and at LABs 
\citep[e.g.][]{Arrigoni2015a,Arrigoni2015b}, and only compact detections
of these lines have been found 
\citep[e.g.][]{Dey2005,Christensen2006,Prescott2009,Prescott2013,Scarlata2009,Yang2010}.
The emission in \ion{C}{IV} and \ion{He}{II} is expected to be 
fainter than the \Lya emission on large scales in all the aforementioned scenarios
\citep{Yang2006,Arrigoni2015a,Arrigoni2015b,Cabot2016}, and consequently only the
brightest compact emission can currently be detected given the surface brightness 
limits achieved so far.
We note that \citet{Prescott2012b} argue that their non-detection of \ion{He}{II}
in narrowband imaging suggests extended emission of \ion{He}{II} around a LAB.

\begin{figure*}
  \centering
  \includegraphics[width = 1.0\textwidth]{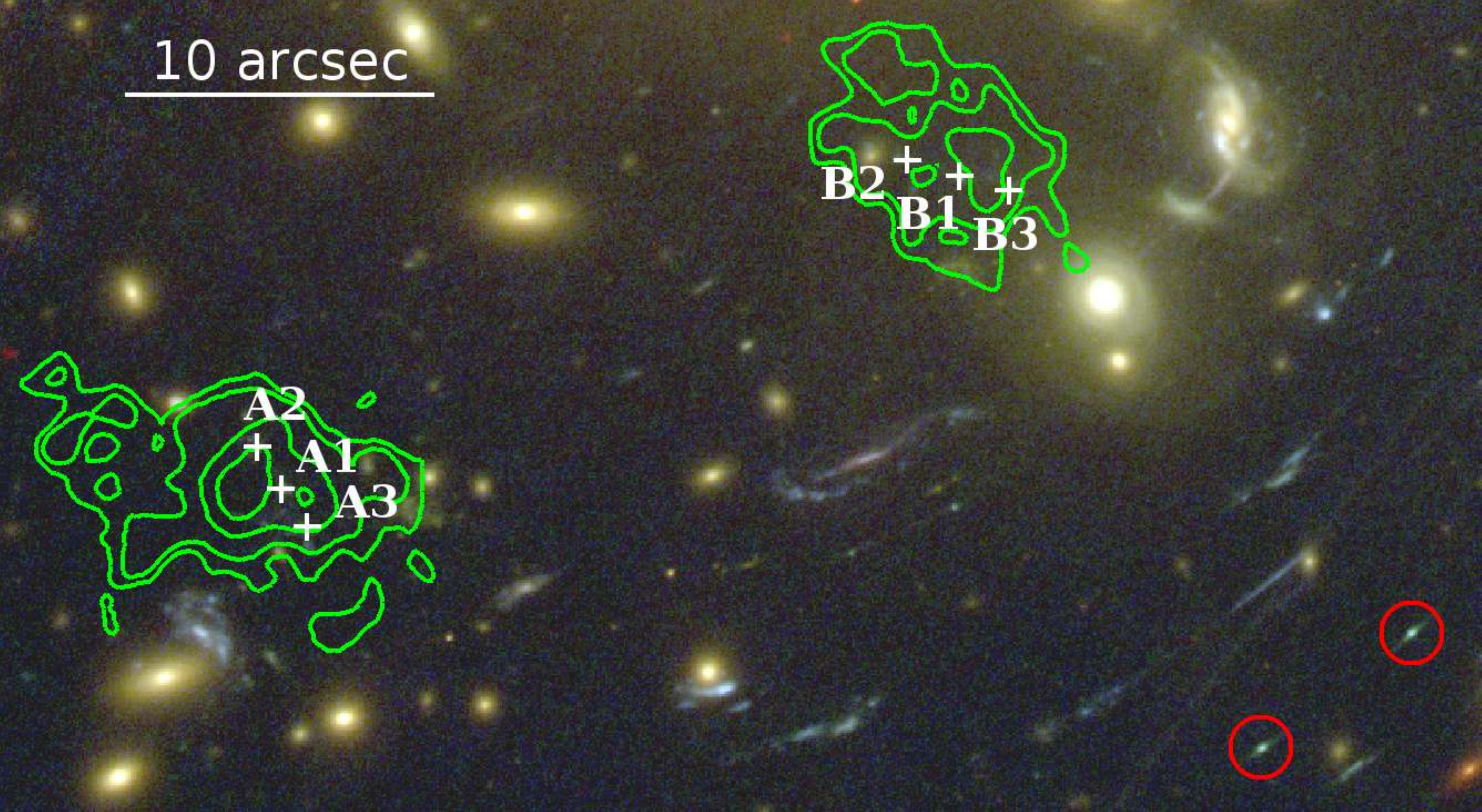}
  \caption{HFF colour composite image of the south-west region of AS1063 overlaid with \Lya emission isocontours. The white crosses indicate the positions of the three compact sources related to the LAB (as in Fig. \ref{fig:images}) and the red circles the multiple images of another galaxy located at the same redshift, studied in detail in \citet{Vanzella2016b}.}
  \label{fig:lya_cont}
\end{figure*}

In this work, we report the discovery of one of the first multiply lensed LABs. Located
behind the Hubble Frontier Fields \citep[HFF; P.I.: J. Lotz, see][]{2016arXiv160506567L} cluster Abell S1063 (AS1063), it is one of the intrinsically-lowest luminosity LABs found to 
date, and the faintest to also show \ion{C}{IV} and \ion{He}{II} emission. We will
demonstrate that the properties of this LAB are consistent with resonant scattering of an 
embedded star-forming source, shedding light on the origin of faint LABs. 
The layout of this paper is as follows.
We present the data and method in Sect. \ref{sec:methodology}, our results in Sect.
\ref{sec:lensed_lya}, and our conclusions in Sect. \ref{sec:conclusions}. Throughout this 
work we use the standard flat $\Lambda$CDM cosmology, with 
$H_0~=~70~$\kms~Mpc$^{-1}$ and $\Omega_M~=~0.3$,
we use a Chabrier initial mass function (IMF), and all magnitudes refer to the AB system.

\section{Methodology}
\label{sec:methodology}

\subsection{Data}
\label{sec:data}

We observed AS1063 with the Multi Unit Spectroscopic Explorer \citep[MUSE,][]{Bacon2010} mounted 
on the Very Large Telescope (VLT) at Paranal for a total of three hours of exposure 
time. The observations were collected
as part of the Science Verification, and the data are presented in 
\citet{Karman2015a}. We refer the reader to that work for a 
description of the data reduction, of which provide a quick overview here. The data
was reduced with the standard pipeline version 0.18.2 and we verified that 
using a later version does not make a significant difference. 
The final datacube covers a field of 1 arcmin$^2$ in the south-west half
of AS1063, and spans a spectral range of 4750 through 9350~\AA. The spectral
resolution of the MUSE instrument increases from $R\approx1700$ at the blue to
$R\approx3500$ at $\sim$9000 \AA, which is enough to spectrally resolve the shape 
of \Lya\ emission. The seeing during the observations varied between
1.0\arcsec and 1.1\arcsec measured in white light, which corresponds to a FHWM of five to six pixels.

We re-examined the datacube after the first publication, inspecting specifically
those positions with photometrically determined multiple images. We reported
in \citet{Caminha2015} the redshifts of two new multiply lensed sources, of which one is an extended
\Lya source, i.e. the LAB studied in this work, see Fig. \ref{fig:lya_cont}. We extracted spectra at the positions of the multiple images 
within our field of view, $\alpha$~=~22:48:44.98; $\delta$~=~$-$44:32:19.1 and 
$\alpha$~=~22:48:42.91; $\delta$~=~$-$44:32:09.0, with differently sized apertures,
to constrain the size and extent of this emission.

\begin{figure}
  \includegraphics[clip, width = .328\columnwidth]{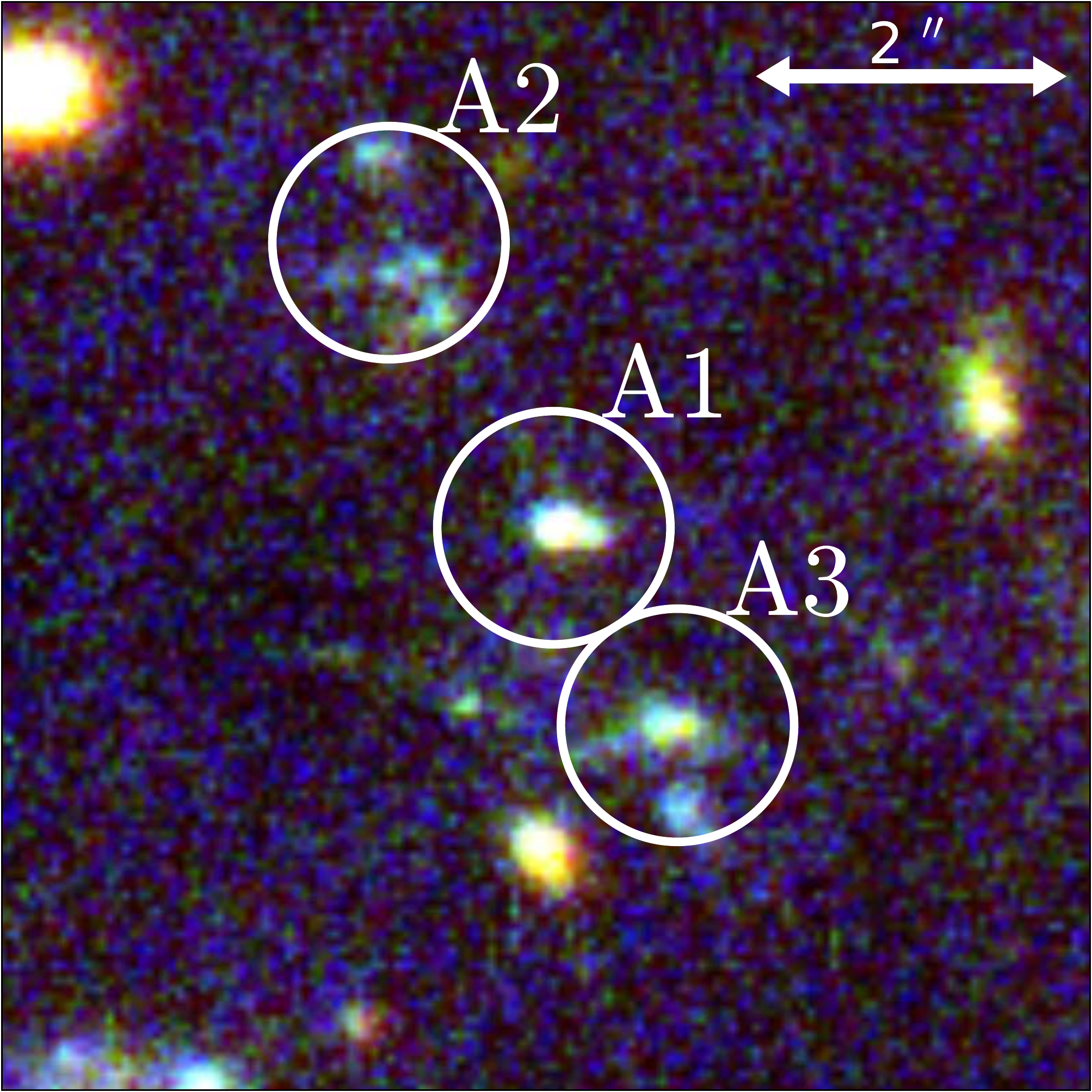}
  \includegraphics[width = .328\columnwidth]{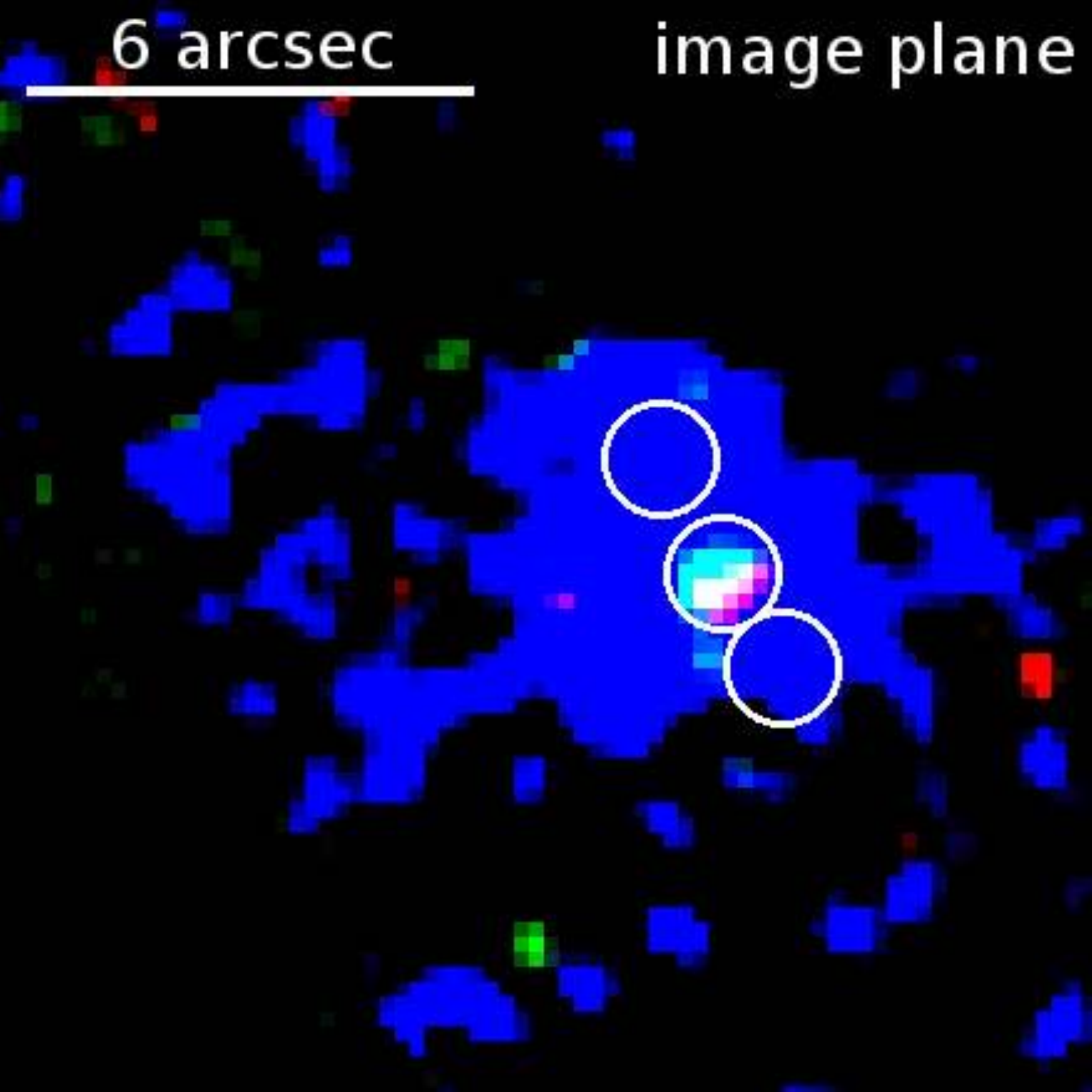}
  \includegraphics[width = .328\columnwidth]{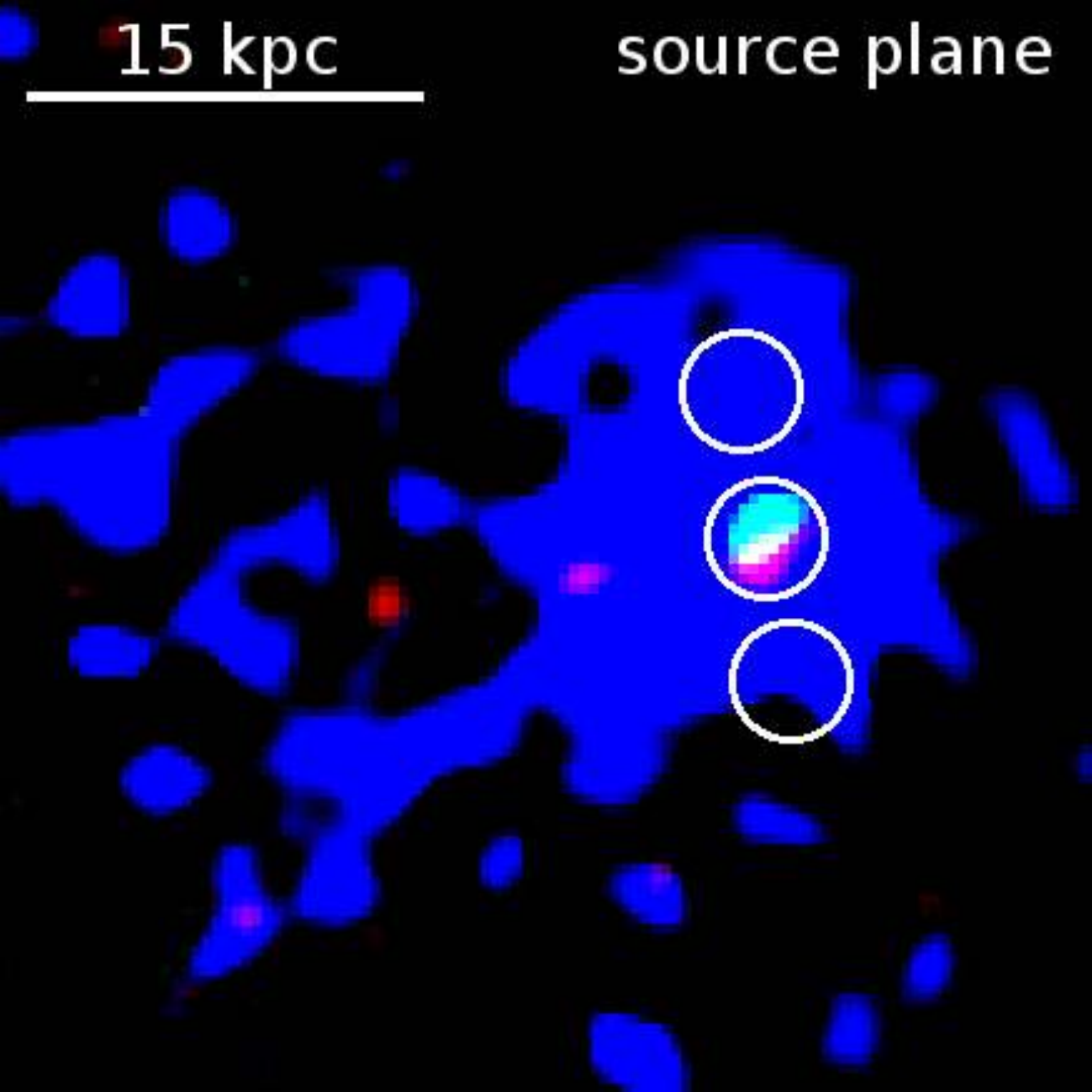}
  
  \includegraphics[clip, width = .328\columnwidth]{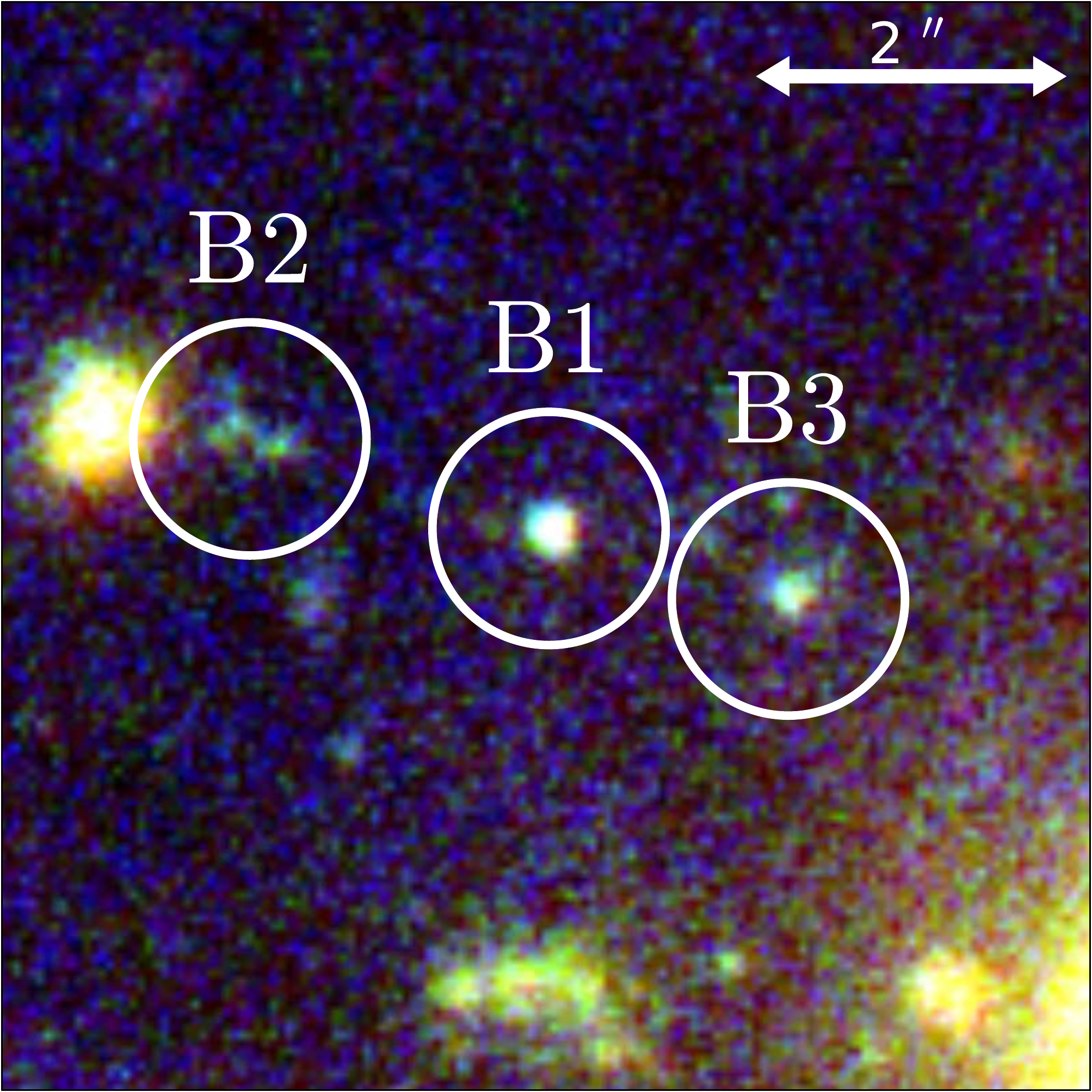}
  \includegraphics[width = .328\columnwidth]{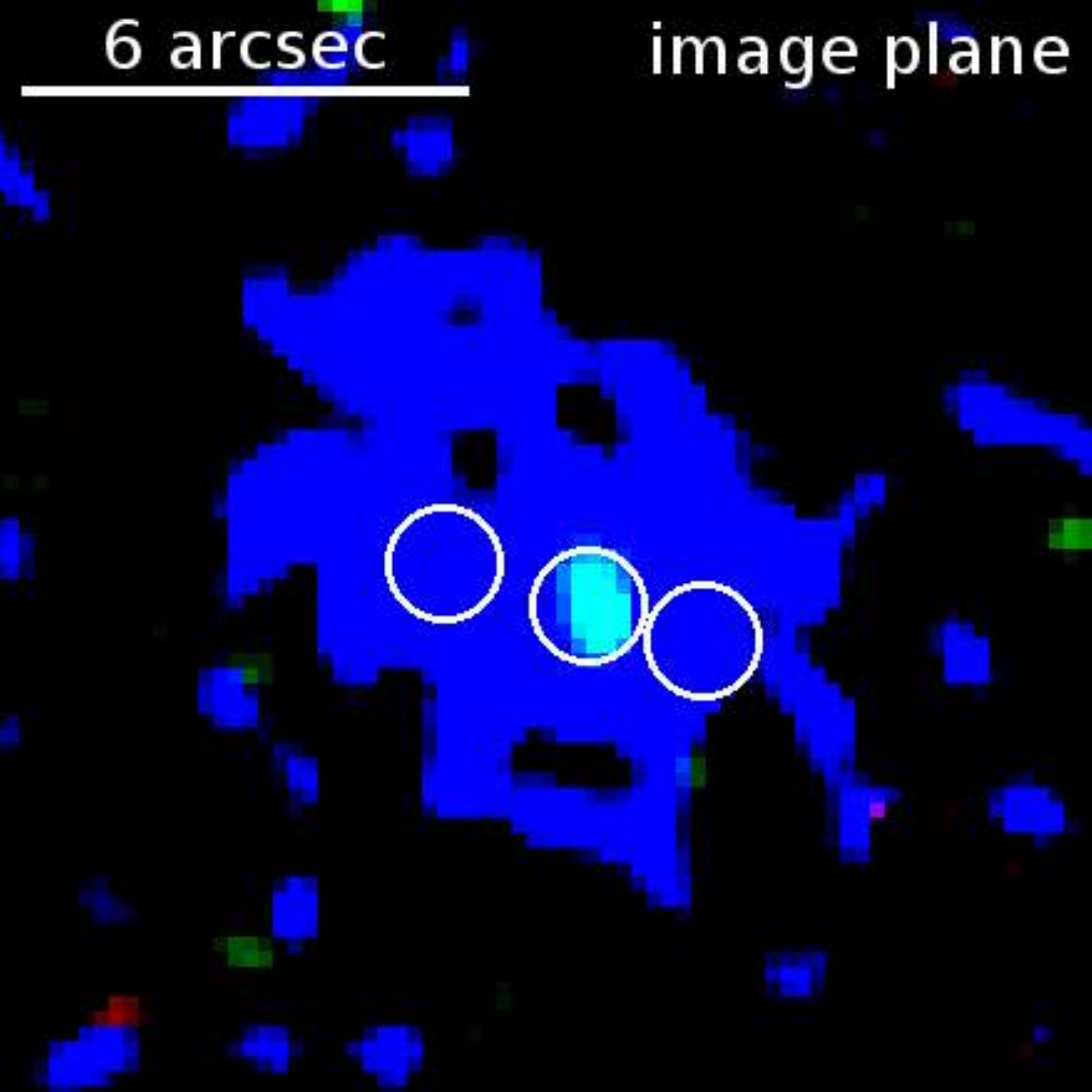}
  \includegraphics[width = .328\columnwidth]{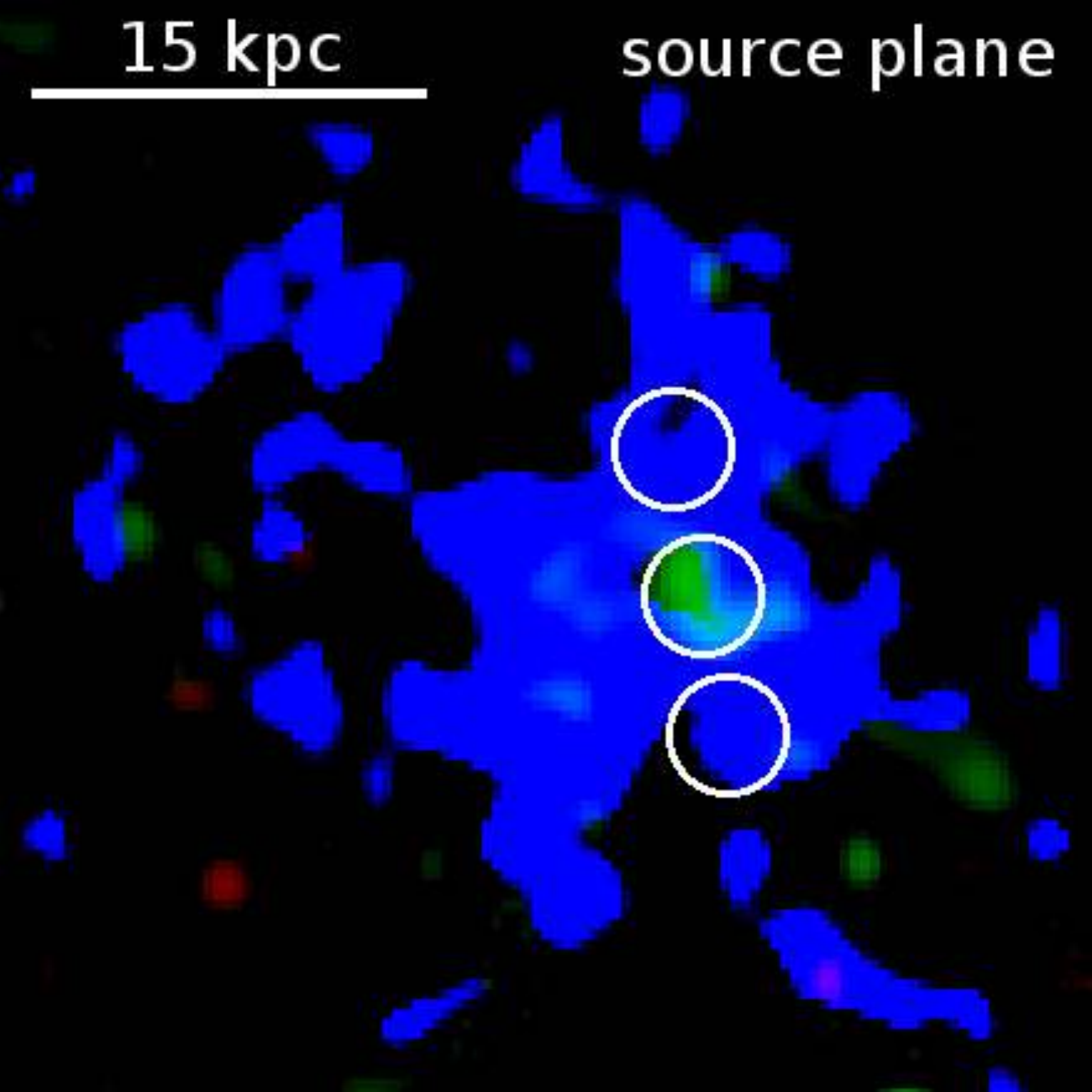}

  \includegraphics[clip, width = .328\columnwidth]{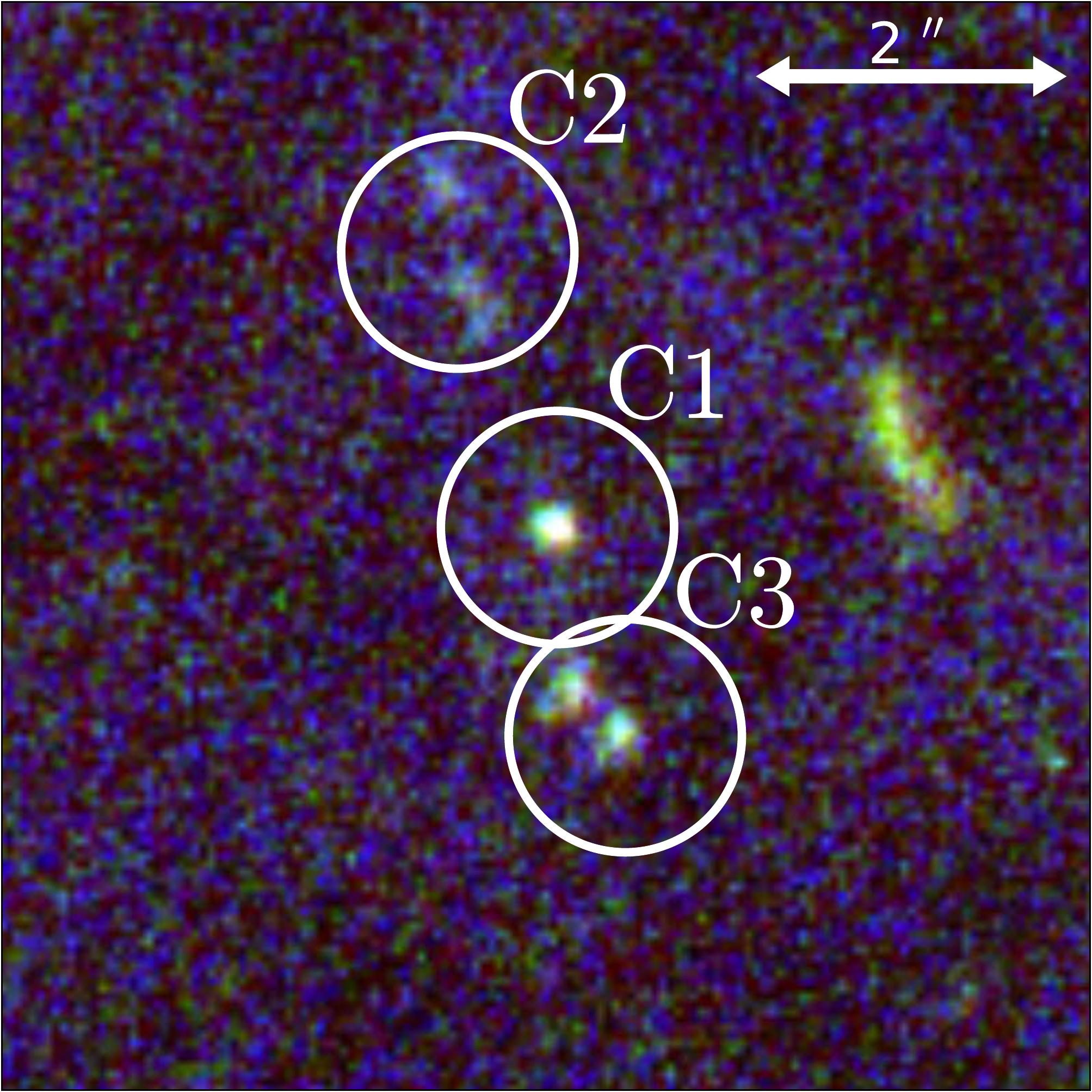}

  \caption{Left panels: details of the LAB multiple images in the HFF colour composite image using the \emph{HST}/ACS F435W, F606W and F814W filters. The circles have $0\arcsec.75$ radius showing the compact source in the centre (ID 1) and the filamentary substructure (IDs 2 and 3). Central panels: colour composite images using the \Lyac \ion{C}{IV} and \ion{He}{II} emissions (blue, green and red channels, respectively) of the two LAB in the MUSE footprint, with a smooth kernel of $1\arcsec$. Right panels: projected emissions onto the source plane. Note that the multiple images C are outside of our MUSE observations, and therefore we cannot show the corresponding Lya image reconstruction.}
  \label{fig:images}
\end{figure}

\subsection{Strong lens model}
\label{sec:strong_lens_model}
This LAB is multiply lensed into three multiple images (A, B and C, see Fig. \ref{fig:images}) and highly distorted and magnified by the gravitational potential of the foreground galaxy cluster AS1063.
To compute the intrinsic properties of the LAB we corrected the observations from the strong lensing effect.
We used an updated model of our previous strong lensing model ID F1-5th, presented in \citet{Caminha2015} to map the quantities between the source and image (observed) planes.
We improved this model by fixing the redshift of the Ly$\alpha$ blob and we included two extra multiple images (structures of the Ly$\alpha$ blob) identified in the HFF images.
We show these new multiple images in Fig.\ref{fig:images}, identified as A2, A3, B2, B3, C2 and C3, see Table \ref{tab:coords} for their coordinates.
The multiple images A1, B1 and C1 correspond to the first identified compact source in the centre of the LAB considered in \citet{Caminha2015}{, and have {\em Hubble Space Telescope} F814W magnitudes of 25.39, 25.81, and 26.38, respectively \citep[see][]{2016arXiv160601471K}.}
The morphology, parity and colours of these substructures in all the three multiple images (A, B and C) are in full agreement with basic gravitational lensing theory, which ensures these substructures are at the same redshift.
In this model we used only spectroscopically confirmed families, totalling 30 multiple images belonging to ten multiple image families of which eight are at different redshifts.
The mean offset between the observed and model predicted positions of the images (A1, B1, C1), (A2, B2, C2) and (A3, B3, C3), is $\approx 0\arcsec.1$, ensuring that our model reproduces the intrinsic properties of these sources very well.
The multiple images A and B are in the MUSE footprint and the mean magnification factors over the extended region of the \Lya emission are $6.1 \pm 0.5$ and $5.0 \pm 0.4$, respectively. This corresponds to a delensed F814W magnitude of 27.36 and 27.56 for Images A1 and B1, respectively.

We used our strong lensing model to map the emission in the image plane onto the source plane, central and right panels of Fig. \ref{fig:images}, respectively.
In the case of Image B, the presence of nearby bright cluster members affects the reconstruction on the source plane significantly, making it more noisy than the reconstruction of Image A.
We note that the observational seeing, not taken into account in this projection, might affect the shape of small scale features on the source plane.
In both reconstructions in the source plane the extended nature of the \Lya emission remains and extends up to $\approx 33$ kpc.

\section{The multiply lensed Lyman-$\alpha$ blob}
\label{sec:lensed_lya}

\begin{figure*}
  \centering
  \includegraphics[width = \textwidth]{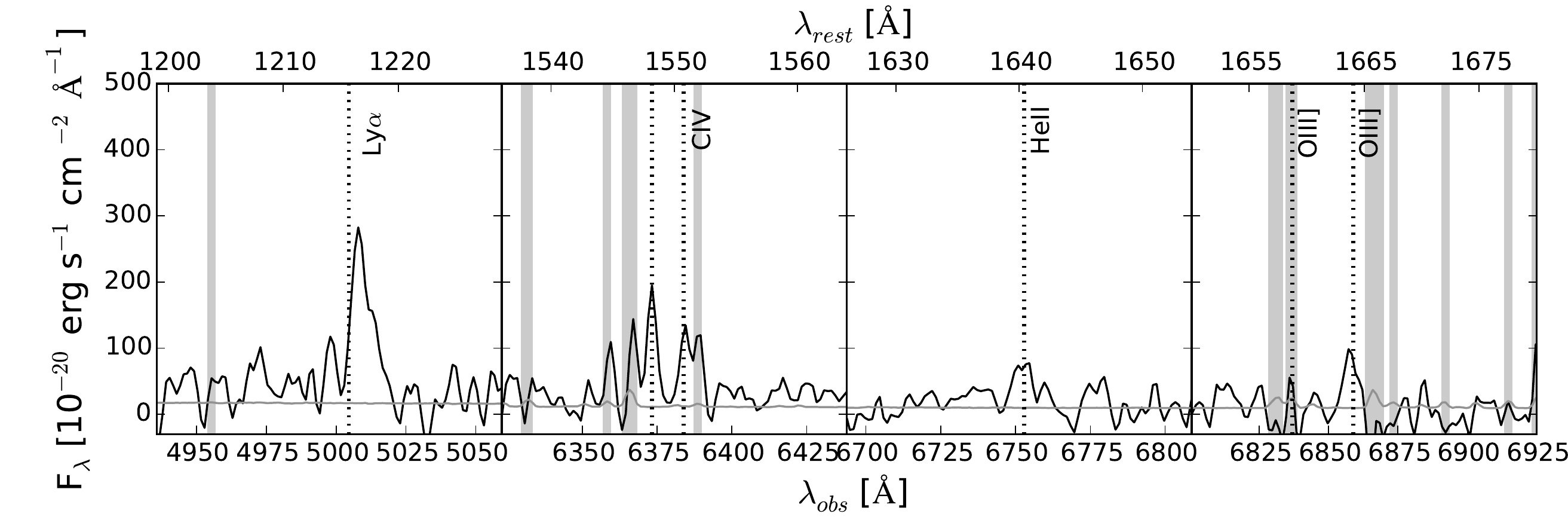}
  \caption{Spectrum of the LAB Image A1 extracted with a circular aperture with 1$\arcsec$ radius, zoomed in on \Lyac \ion{C}{IV}, and \ion{He}{II} from left to right. The black line shows the spectrum, the grey line shows the typical error, the grey vertical bands are located at wavelengths with significant sky interference, and the dotted vertical lines show the location of the restframe emission lines.}
  \label{fig:spectrum}
\end{figure*}

We found several emission lines in the spectra of compact sources A1 and B1 at the centre of the LAB, which we identified as \Lyac \ion{C}{IV}$\lambda\lambda$1548,1551~\AA, \ion{He}{II}$\lambda1640$~\AA, and \ion{O}{III}]1666~\AA, determining the redshift to be $z=3.117$. We show the \Lyac \ion{C}{IV}, \ion{He}{II}, and \ion{O}{III}] emission lines of spectra of Image A1 in Fig. \ref{fig:spectrum}. 

We used the integral field capabilities of MUSE to study the spatial profile of \Lya, and found a significantly extended emission area, see Fig. \ref{fig:lya_cont} and the central and right panels of Fig. \ref{fig:images}.
In Fig. \ref{fig:images}, the \Lya emission of each multiple image appears to be extended up to $\approx 10 \arcsec$ on the image plane and $\approx 33$ kpc on the source plane. In Fig. \ref{fig:lya_cont}, we show the contours of the broad spatial \Lya emission, overplotted on a colour composite image constructed from the HFF images.
From this image, we identified three compact sources embedded within the \Lya emission, which are shown in detail in the left panels of Fig. \ref{fig:images}.
One of these compact sources is very bright and lies at the centre of the emission while the other two present a filamentary structure.
These two other sources are seen around all three multiple images, meaning that they are at a similar redshift. They are close enough to the central galaxy that they fall within the projected area of the \Lya emission. We note that the contours corresponding to the highest flux levels are located around images A1 and B1, rather than on top of it.

\begin{table}[!ht]
\centering
\caption{Positions of the multiple images associated to the LAB (IDs A, B and C) and the positions (IDs N) throughout the nebula for which we extracted the spectra. \label{tab:coords}}
\begin{tabular}{l c c c c l} \hline \hline
ID & $\alpha$ & $\delta$ \\ \hline
A1 & 22:48:44.98 & $-$44:32:19.1 \\ 
A2 & 22:48:45.05 & $-$44:32:17.9 \\ 
A3 & 22:48:44.92 & $-$44:32:20.5 \\ 
B1 & 22:48:42.91 & $-$44:32:09.0 \\
B2 & 22:48:43.09 & $-$44:32:08.6 \\
B3 & 22:48:42.78 & $-$44:32:09.6 \\
C1 & 22:48:40.96 & $-$44:31:19.5 \\
C2 & 22:48:41.01 & $-$44:31:18.0 \\
C3 & 22:48:40.92 & $-$44:31:21.0 \\  
N1 & 22:48:45.08 & $-$44:32:19.8 \\ 
N2 & 22:48:44.88 & $-$44:32:18.8 \\ 
N3 & 22:48:45.21 & $-$44:32:18.9 \\ 
\hline \hline
\end{tabular}
\label{tab:extract_pos}
\end{table}

\begin{figure}
  \centering
  \includegraphics[width = 1.0\columnwidth]{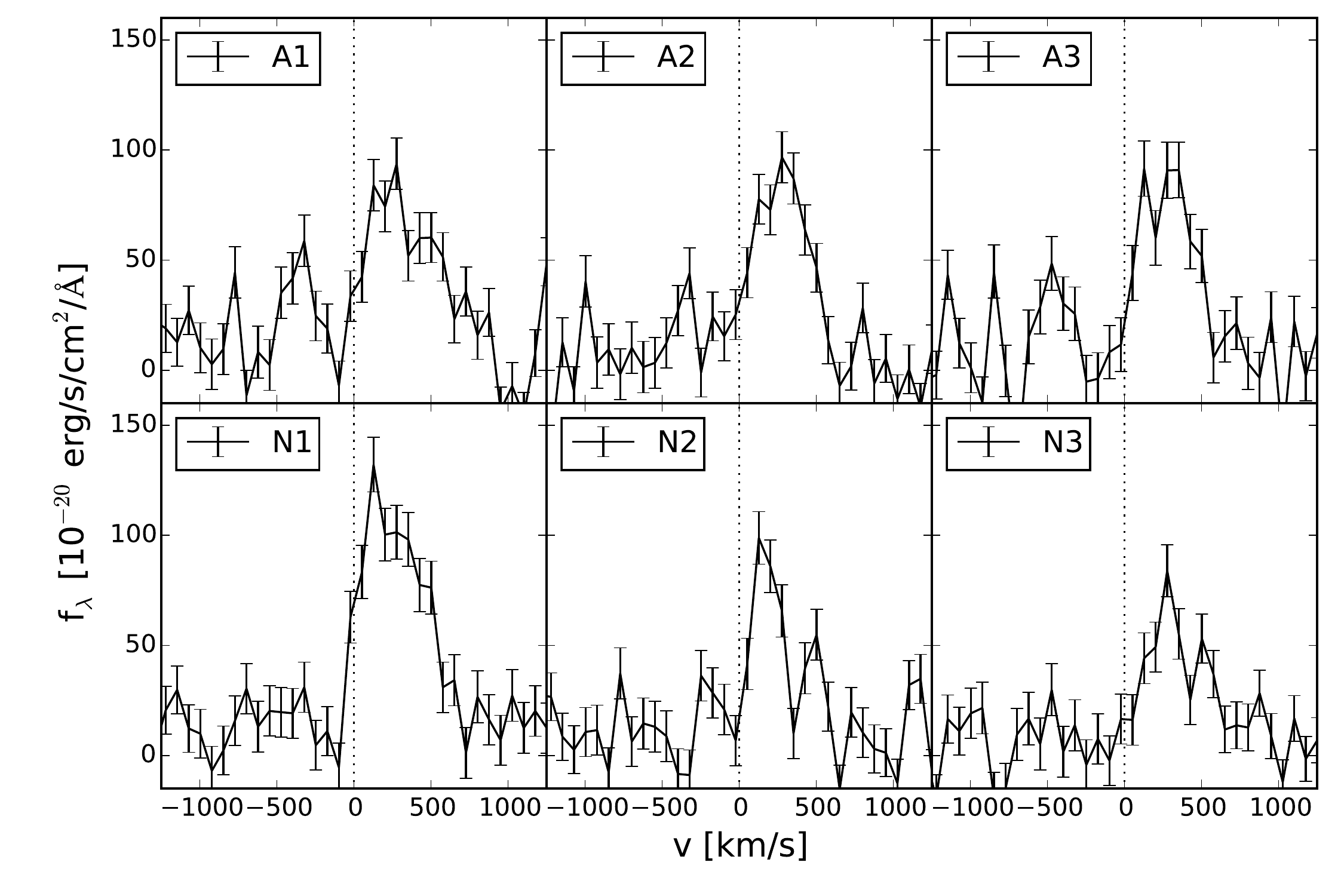}
  \caption{\Lya profile at different positions of the nebula. The top row shows the \Lya line at the HST counterparts A1, A2, and A3, while three positions throughout the nebula are shown by N1, N2, and N3. The spectra have been summed within a four pixel apperture radius and the coordinates are shown in Table \ref{tab:extract_pos}.\label{fig:shapes}}
\end{figure}

The shape of the \Lya profile can give us insight into the properties of the gas surrounding this galaxy \citep[e.g.][]{Verhamme2008,Gronke2015}. At the position of the brightest counterpart, the \Lya line has a typical asymmetric profile with a red tail and a small blue second peak separated by $\sim600$~km~s$^{-1}$ from the red peak which is offset by $\sim200$~km~s$^{-1}$ from the wavelength predicted from the ultraviolet (UV) emission lines. This line profile suggests a shell of gas flowing out of the galaxy at moderate velocity.
We extracted spectra at the positions of these companions (IDs A2 and A3) and the central compact object (ID A1) finding no significant difference in the \Lya profile, see the top row of Fig. \ref{fig:shapes}, or any emission line at other wavelengths. This means that these two sources are not bright enough in \Lya to be detected or distinguished from the \Lya nebula by our observations.
Moreover, the \Lya profile at different positions of the nebula is very homogeneous, with no significant velocity offsets from the central position. This was also found by previous studies \citep{Swinbank2007,Patricio2015}, and indicates that the kinematic conditions of the nebula are similar in all directions.
It is interesting to note that a deficit of the \Lya emission is seen close to the position of the bright central source, where the contours present a `valley'. This is also illustrated in Fig. \ref{fig:shapes}, where we compare the \Lya line at the bright central object A1 to other positions throughout the nebula. In addition to Image A1, we show the spectra at the other detected objects A2 and A3, and three positions without any optical counterpart. We chose N1 to be situated at the peak of \Lya flux, that is offset from A1, while positions N2 and N3 are chosen to sample the whole nebula. The peak of \Lya flux is less at A1 than at N1, and comparable to the other regions N2 and N3.
The coordinates of these positions are shown in Table \ref{tab:extract_pos}.
We showed in the foregoing analysis that this is most probably due to gas \Lya transfer processes and not to the presence of dust.

The \ion{C}{IV} doublet is clearly detected, and shows a narrow line profile, while the \ion{He}{II} line is detected at $\sim4.8\sigma$ and is broader. The width of the \ion{C}{IV} lines remains unresolved, meaning that the width is equal to the instrumental resolution, while the signal to noise of the \ion{He}{II} lines is too low to reliably measure its width. Although broad emission lines would indicate AGN activity \citep[but see][for broad emission lines in Wolf-Rayet stars]{Crowther2006}, narrow emission lines can be produced by both AGN and SF.

\begin{figure*}
  \centering
   \includegraphics[width = 1.0\columnwidth]{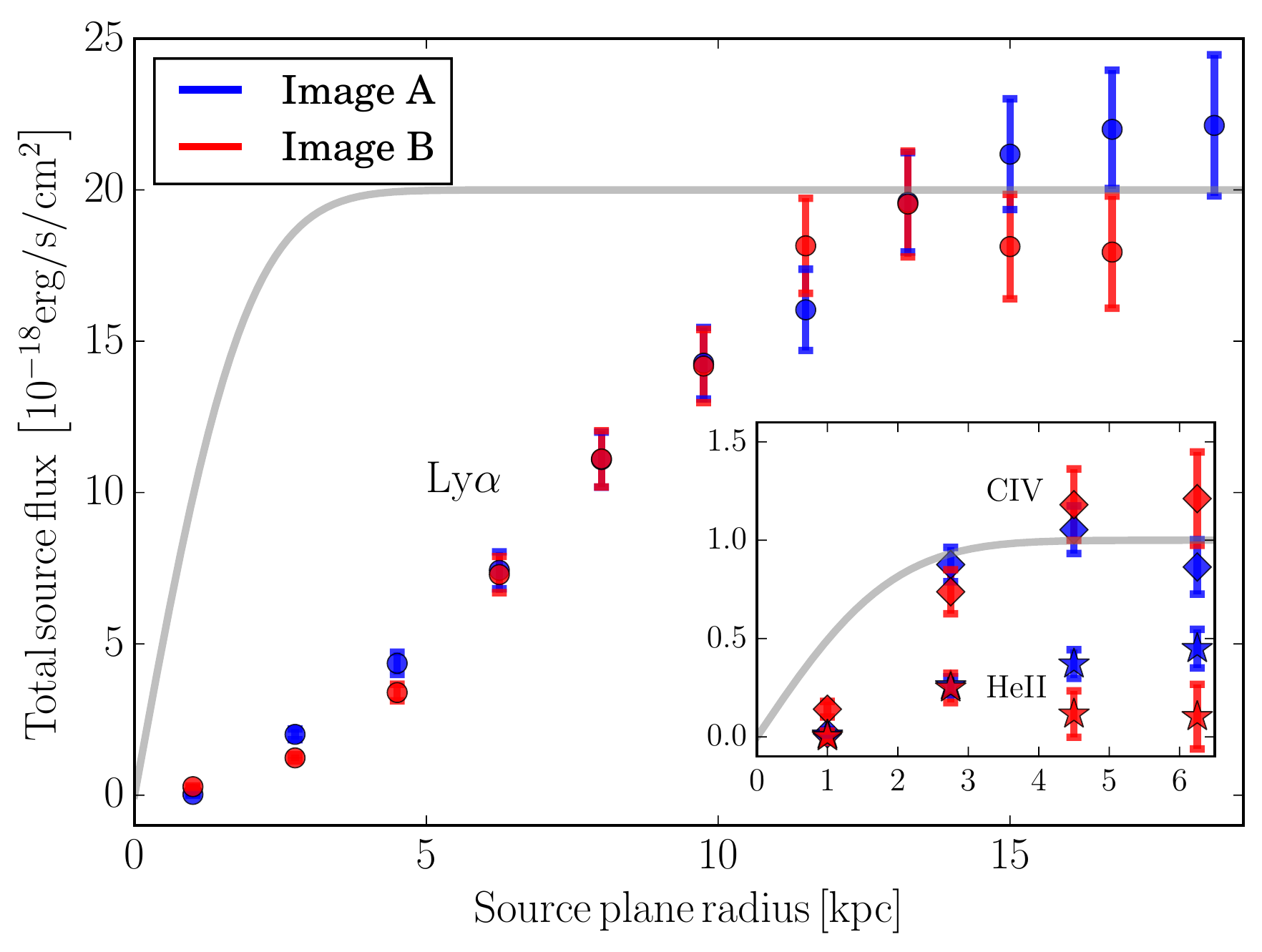}
   \includegraphics[width = 1.0\columnwidth]{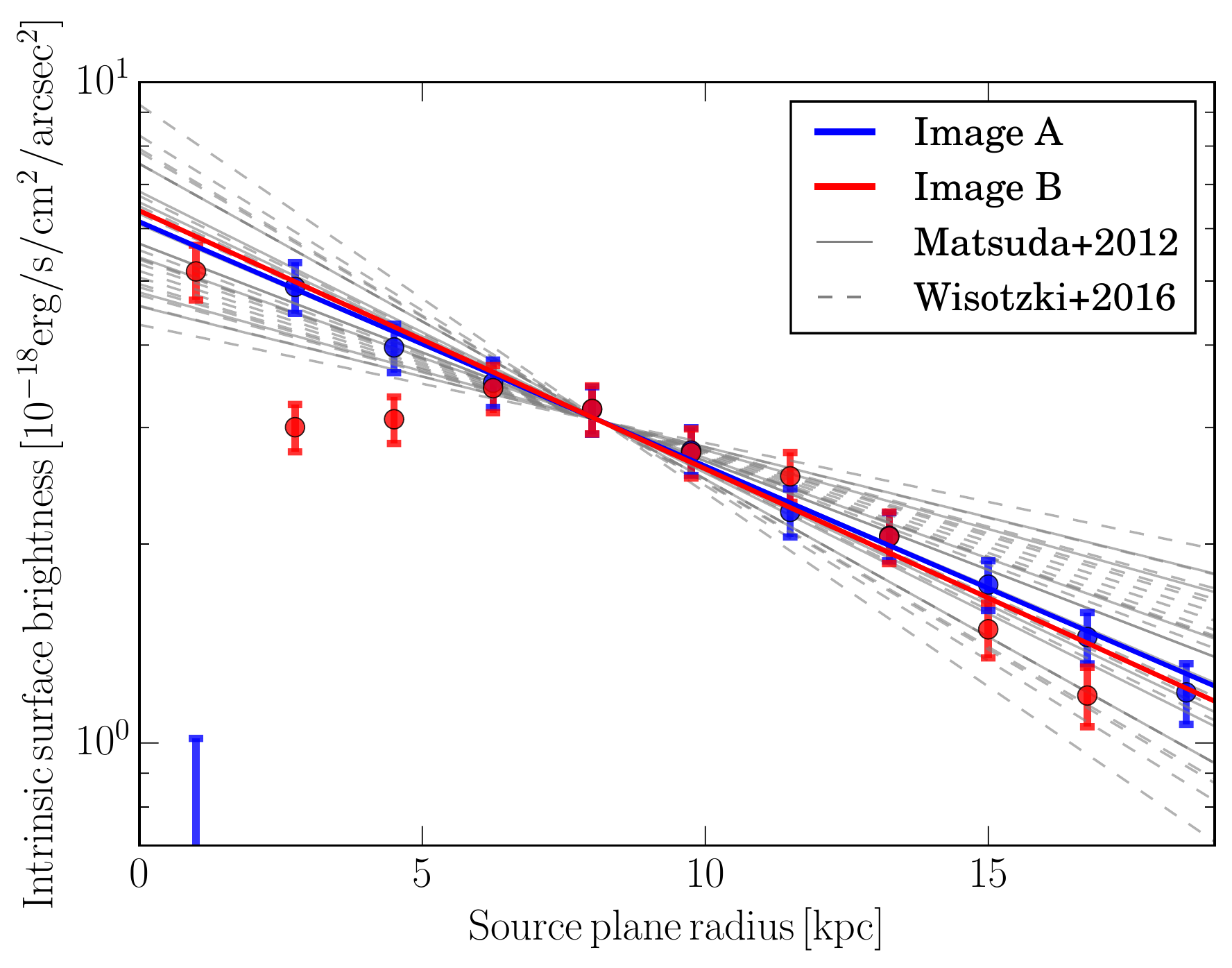}
  \caption{Left panel: total flux on the source plane of the two multiple images of the LAB within the MUSE footprint, i.e. A1 and B1. The circles are the \Lya emission. In the inset plot we show the \ion{C}{IV} and \ion{He}{II} emissions, diamonds and stars respectively (the units are the same as in the main plot). The grey lines show the theoretical shape on the source plane of a Gaussian emission with observed scale of $1\arcsec.1$ and arbitrary total flux. Right panel: intrinsic \Lya surface brightness. The red and blue lines indicate the best fit of an exponential function (considering the points $r>5$ kpc, see text for details). The grey solid and dashed lines are the results from \citet{2012MNRAS.425..878M} and \citet{2016A&A...587A..98W}, respectively, renormalized to $r=8$ kpc.}
  \label{fig:la_profile}
\end{figure*}

To investigate the powering source of the LAB, we studied the extent of the different emission lines.
In the left panel of Fig. \ref{fig:la_profile}, we show the intrinsic cumulative flux profile as a function of the radius on the source plane for the two images integrating in the range of each emission and subtracting the continuum.
The \Lyac \ion{C}{IV} and \ion{He}{II} emission fluxes are the circles, diamonds and stars, respectively.
The flux centred on Image B1 is rescaled to account for the flux ratio anomalies that can be caused by small mass substructures not included in the lens model \citep{1991ApJ...373..354K,2007MNRAS.378..109M}.
To highlight the extended nature of the emission, the grey lines show the shape of a Gaussian point source in similar seeing conditions.

It is clear from these figures that the \Lya emission is extended, while the \ion{C}{IV} line is significantly more compact than the \Lya and cospatial with Source 1. This suggests that the nebula is not powered by photoionization or shocks. The extent of the \ion{He}{II} line is more difficult to determine, as the flux density is low. Therefore, we cannot distinguish between gravitational cooling or resonant scattering, although the presence of a compact ionizing source suggests the latter. 

In the right panel of Fig. \ref{fig:la_profile}, we show the \Lya surface brightness of sources A1 and B1 as a function of the radius on the source plane.
The best fit, considering the scale of the surface brightness with the radius ($r$) given by $S(r) = C \exp (-r/r_n )$, where $C$ is a normalization and $r_n$ gives the slope of the curve, is shown in the Figure.
To avoid systematics in the fitting we excluded the points with $r<5$~kpc for which the area is small and the measured flux more noisier and the observational seeing might contribute significantly in the overall slope.
The best fitting values of $r_n$ for the Images A1 and B1 are $(11.8 \pm 0.4)$~kpc and $(11.1 \pm 1.3)$~kpc, respectively.
These values show good agreement with the measured slopes in \citet{2012MNRAS.425..878M} and \citet{2016A&A...587A..98W}, also shown in the Figure, although our \Lya surface brightness profile falls in the fainter range of these two samples and we cannot extend our fit to large radii due to the low surface brightness.

\begin{figure}
  \centering
   \includegraphics[width = 1.0\columnwidth]{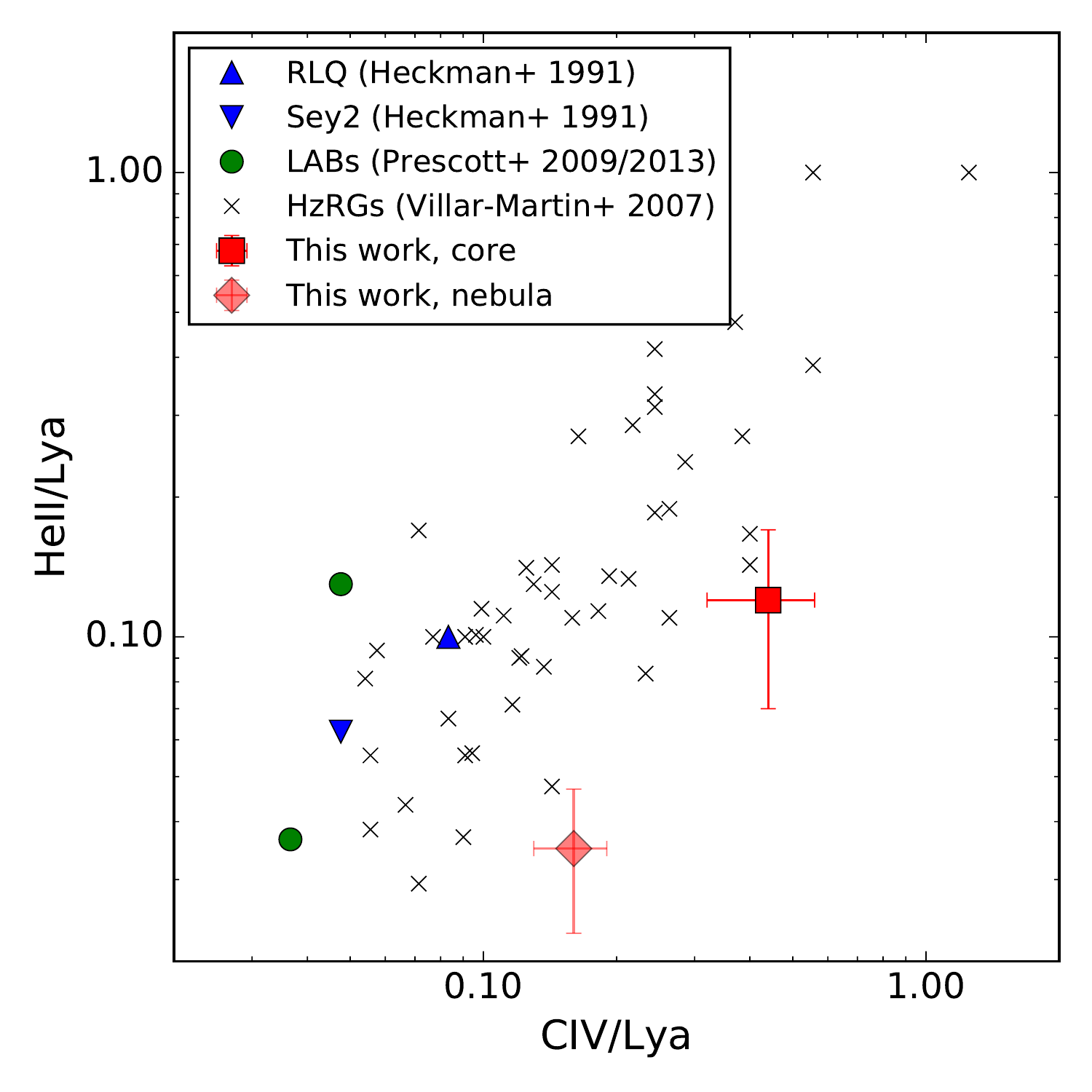}
  \caption{Flux ratio of \ion{He}{II}/\Lya versus \ion{C}{IV}/\Lyap The red square corresponds to the central region of the \Lya emission, while the red semi-transparent diamond corresponds to the full nebula. We compare our results to the values found in other LABs \citep[green circles][]{Prescott2009,Prescott2013}, HzRGs \citep[black crosses][]{Villar2007}, and radio loud quasars and Seyfert 2 hosts \citep[blue triangles][]{Heckman1991a}.}
  \label{fig:CIV_HeII}
\end{figure}

The ratio of \Lya to \ion{C}{IV} or \ion{He}{II} is often used to determine the ionizing source of LABs, even though effects on the \Lya emission like dust absorption could bias these ratios. In Fig. \ref{fig:CIV_HeII} we show our data together with the detections reported in the literature for \Lya nebulae at comparable redshift \citep{Heckman1991a,Villar2007,Prescott2009,Prescott2013}. In particular, we measured \ion{C}{IV}/\Lya = 0.44$\pm0.12$ and \ion{He}{II}/\Lya= $0.12\pm0.05$ from a central region of $0\arcsec.5$, and \ion{C}{IV}/\Lya = 0.16$\pm0.03$ and \ion{He}{II}/\Lya = $0.03\pm0.01$ over the complete nebula, see Fig. \ref{fig:CIV_HeII}. We see that the ratio of \ion{C}{IV}/\Lya is high for its given \ion{He}{II}/\Lyac and that the other LABs have a much lower \ion{C}{IV}/\Lya ratio, while \ion{He}{II}/\Lya is similar. This specific line ratio in the inner parts is in agreement with an AGN, while the ratio of the total nebula indicates SF as the main source of the \Lya emission \citep[][]{Binette2006,Villar2007,Swinbank2007}, making it necessary the use of the \ion{O}{III} emission to discriminate between these two scenarios. This shows the importance of integral field spectroscopy when studying ionization sources.

Recently, \citet{Feltre2015} published predictions of UV line ratios for AGNe and SF galaxies. Unfortunately, one of the main lines that they discuss to separate dominant AGN from SF galaxies and determine ionization properties is \ion{C}{III}], which we did not detect due to skylines. Alternatively, they show that the ratio of \ion{O}{III}]/\ion{He}{II} is able to separate AGNe and SF galaxies for most values. We measured Log(\ion{O}{III}]/\ion{He}{II})=$0.35\pm0.23$  in the central source, which indicates that SF is the dominant ionizing source in the core of the LAB. All other possible line ratios we were able to measure are consistent with SF being the dominant factor. To exclude possible AGN dominance, we ran a set of CLOUDY \citep[last described by][]{Ferland2013} models following \citet{Arrigoni2015a}. We found that AGN-powered nebula have -5.34$<$Log(\ion{O}{III}]/\ion{He}{II})$<$-1.43, which is incompatible with the ratio measured for the central region discussed here.

\begin{figure}
  \centering
  \includegraphics[width = 1.0\columnwidth]{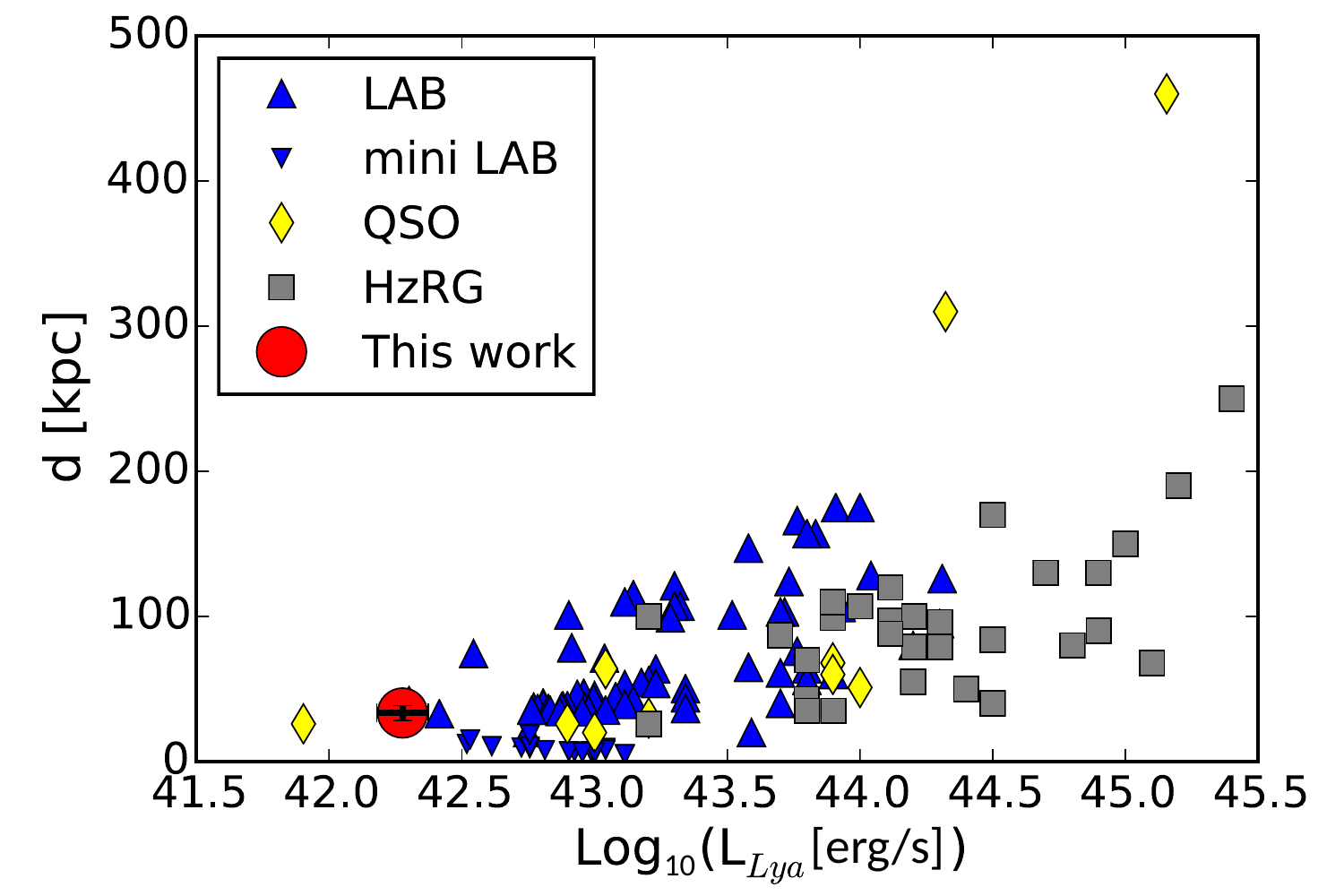}
  \caption{Luminosity and diameter of our target compared to literature values. The LAB sample is taken from \citet{Matsuda2004,Matsuda2011,Ouchi2009,Prescott2013}, and \citet{Patricio2015}, the mini-LABs ($d<30$~kpc) are from 
  \citet{Saito2006,Saito2008}, the HzRGs are from \citet{Heckman1991a} and \citet{Swinbank2015}, and the QSOs are taken from
  \citet{Christensen2006,North2012,Cantalupo2014}, and \citet{Hennawi2015}.}
  \label{fig:luminosity}
\end{figure}

We compare the intrinsic luminosity of this source to other LABs and nebulas reported in the literature in Fig. \ref{fig:luminosity}. Clearly, the LAB studied here is one of the faintest nebula observed to date, and the faintest for which other emission lines are detected. We used $L_{\Lya}$ of the complete nebula to estimate the SF rate (SFR), assuming case B recombination and the conversion of H$\alpha$ to SFR from \citet{Kennicutt1998}, and found \nobreak SFR$_{\Lyan}\approx$~1.7~\Msun~yr$^{-1}$. We note that case B recombination is not occuring in the nebula, but only in the central star forming region. We assume that the escaping \Lya photons are then scattered without significant absorption through the nebula to cause the extended emission.

We performed spectral energy distribution fitting on the broadband photometry
from 0.4 through 2.4 $\mu$m to determine the stellar mass, stellar age, 
and SFR \citep[see][]{2016arXiv160601471K}. We only considered images A1 and C1, as the 
photometry of Image B is contaminated by nearby galaxies in several bands.
We find a stellar mass of ${\rm Log}(M_\star) = 7.94^{+0.20}_{-0.16}$ 
($7.85^{+0.20}_{-0.12}$) for images A (C).
The SFR and stellar age are less constrained, with best fit values of
SFR = 2$^{+18}_{-0.4}$ (2.5$^{+250}_{-1.5}$)$M_\odot$ yr$^{-1}$ and 
Log(age/yr) = 6.6 $^{+1.4}_{-0.1}$ (5.8$^{+2.2}_{-0.4}$), finding good 
agreement between SFR$_{\Lyan}$ and SFR$_{\rm SED}$. The galaxy is best fit by
dust-poor templates, with {\it E(B-V)}$\leq$0.2, showing that the central \Lya extinction
is more likely due to transfer processes than dust absorption. We used the best-fitting SED to estimate
the continuum flux, and derived an equivalent width of 110 \AA\ for \Lya,
which can be produced by SF \citep[e.g.][]{Charlot1993}. From the best-fitting SED, we measure
a UV continuum slope $\beta$= $-2.09^{+0.06}_{-0.10}$ ($-2.18^{+0.08}_{-0.22}$).  A young star-forming
galaxy is in agreement with the UV line ratios, and we showed that it could 
be the source producing the scattered \Lya photons.
Although the SED fitting (UV slope) does not suggest large amounts of dust, we cannot exclude some dust attenuation of the \Lya emission and hence an underestimate of the SFR. We note however that the inclusion of little dust would not affect our conclusions about the nature of the luminosity of the nebula.

From the left panels of Fig. \ref{fig:images} we see that the continuum light of sources A1, B1, and C1 is spatially resolved. 
We measured an effective radius of 0.11\arcsec, corresponding to a physical scale of $\sim$360 pc. Because
most of the light is emitted in this resolved region, most of the light cannot be due to an unresolved
point source, for example an AGN. This further strengthens our conclusion that SF is producing the scattered
\Lya photons, rather than an AGN.

In the MUSE footprint we found two other compact \Lya emitters at the same redshift ($z=3.117$), one is also a multiply lensed galaxy and optically-thin \citep[red circles in Fig. \ref{fig:lya_cont}, and a possible Lyman continuum emitter, see][]{Vanzella2016b}, and the other is reported in Table 2 of \citet{Karman2015a}.
\citet{Vanzella2016b} showed also that the optically-thin source is very young, dust-free and low mass object, belonging to a low-luminosity domain, $L_{1500}\approx0.02L^*_{z=3}$, possibly analogue to the sources responsible for reionization. 
The distance between these three objects on the source plane is $\lesssim 103$ kpc, indicating that the LAB is located in a group of galaxies, and possibly a protocluster.

It is noteworthy that besides the LAB presented here, the only other multiply lensed LAB \citep{Patricio2015} also shows a relatively small and faint \Lya nebula powered by SF, although it is significantly more massive than the LAB discussed here. These faint LABs powered by SF are very similar to those obtained in the local Universe, as found by the Lyman Alpha Reference Sample \citep{Hayes2013}, and indicate that with the new generation of IFUs in combination with gravitational lensing we are now probing the CGM of sub-$L_{*}$ galaxies at $z\sim3$.

\section{Conclusions}
\label{sec:conclusions}

In this work, we presented the discovery of a faint multiply lensed LAB. 
Using MUSE, we obtained spectral and spatial information on the \Lyac 
\ion{C}{IV}, \ion{He}{II}, and \ion{O}{III}] lines, which we used to determine
the powering source of this LAB. We used our gravitational lensing model to 
correct for magnification and to reconstruct intrinsic source properties.

We found an extended \Lya emission region,
with a total luminosity of $1.9\times10^{42}$ erg s$^{-1}$ distributed over a circular
area with a radius of 16.5 kpc on the source plane. The low luminosity makes this \Lya nebula
the second faintest observed to date, and the faintest LAB with detected 
UV emission lines. 

Besides \Lyac we found no other lines with extended 
emission, although the \ion{He}{II} flux density is too low to exclude extended
emission. Unlike \ion{He}{II}, the flux density of the \ion{C}{IV} emission
is high enough to rule out extended emission, which indicates that photoionization or
shocks cannot be the powering mechanism of this LAB. In addition, the deficit
of \Lya at the centre of the LAB hints at resonant scattering rather than
gravitational cooling. The UV-line ratios indicate that the embedded source in 
the LAB is SF rather than an AGN.

Interestingly, we confirmed spectroscopically two other galaxies 
at very close distance from the LAB. This is in agreement with previous
results in the literature that show that LABs reside in overdensities \citep{Yang2009}. With dedicated  
observations in the future, one could quantify the galaxy overdensity and establish the presence of a protocluster behind AS1063.

\begin{acknowledgements}
G.B.C. is supported by the CAPES-ICRANet program grant BEX 13946/13-7.
C.G. acknowledges support by VILLUM FONDEN Young Investigator Programme grant 10123.
We acknowledge financial support from PRIN-INAF 2014 1.05.01.94.02.
This work made use of the CHE cluster, managed and funded by ICRA/CBPF/MCTI, with financial support from FINEP (grant 01.07.0515.00 from CT-INFRA - 01/2006) and FAPERJ (grants E-26/171.206/2006 and E-26/110.516/2012).
This work also made use of data gathered under the ESO programmes 186.A-0798 and 60.A-9345(A).
\end{acknowledgements}

\bibliographystyle{aa}
\bibliography{references}

\end{document}